# A Spatiotemporal Context Definition for Service Adaptation Prediction in a Pervasive Computing Environment

Darine Ameyed
MMS Laboratory, Université du Québec,
École de technologie supérieure
Montréal, Canada

Moeiz Miraoui
Al-Leith computer college,
Umm Al-Qura University
Makkah, Saudi Arabia

Chakib Tadj
MMS Laboratory, Université du Québec,
École de technologie supérieure
Montréal, Canada

*Abstract*— **Pervasive systems refers to context-aware systems that can sense their context, and adapt their behavior accordingly to provide adaptable services. Proactive adaptation of such systems allows changing the service and the context based on prediction. However, the definition of the context is still vague and not suitable to prediction.**

**In this paper we discuss and classify previous definitions of context. Then, we present a new definition which allows pervasive systems to understand and predict their contexts. We analyze the essential lines that fall within the context definition, and propose some scenarios to make it clear our approach.**

## I. INTRODUCTION

Since its introduction by Mark Weiser in 1991, the ubiquitous computing presented progressively a new vision of computer systems, where technology becomes invisible to the user, and allows for the proactive adaptation to changes in context of users and applications. From this vision evolved the current interest in context-awareness, supporting the approach of adapting context-aware services, which should be directly related or even deduced from the contextual information. Hence come the notion of context-aware computing.

Implicitly, a context was defined as the set of information pertaining to an event or a particular situation in any environment or supervising communication between humans.

In computer science, the context is specified in a mobile, smart, and scalable environment. This environment provides the ubiquitous systems [1], whose objective is to allow a transparent human-machine and / or machine-machine interaction.
In order to ensure this interaction, the pervasive system must opt for a context-aware adaptation. Defining and collecting of effective contextual information become crucial, and entails challenges that constantly increase with technological development emphasizing mobility for users, and equipment.

Several basic elements of research, in context-aware computing have been proposed in order to answer some key issues, including the context and context awareness [2], [9]. This might be summarized by questions such as: what's the context? How to make use of a context? How to develop context-aware applications?

There are also many other researches that used the specific context information and the development of context aware applications. Such research had already led to a large number of prototypes and even end-user products [8], [29], [32], and [16].

Most of work on the adaptive services based on context has focused on the current context (i.e. the present context). However, a new trend in research on context-aware adaptive services approaches supports taking into account the temporal evolution of context to predict a future context [26], [32], and [30].

Being aware of the future context allows pervasive computing system (PCS) to choose the most effective strategies to achieve its objectives and therefore to provide an active and fast adaptation to future situations.

The first step towards the development of a context-aware pervasive system is to understand the context and establish its components. This requires a clear description of what a context is, both in the development of logical representations and in the processing, in order to use the context efficiently and easily in applications.

Many attempts have been made to provide a concise definition of context. However, the current definitions are still vague and general. Also all the proposed definitions do not take into account new adaptations based on the prediction strategies.

In this paper, we propose a new definition of context, taking into account the goals such as: predicting the context and context-aware adaptive services.

We try to introduce the context in a model highlighting the spatiotemporal dynamics of context elements in the context-aware proactive adaptation.

Understanding how to use the general context while designing adaptive application can help application designers to determine how this context must be implemented in their applications to meet the use requirement.

The rest of the paper is organized as follows: Section II provides an overview of context definitions in the literature. Section III presents our proposed definition of context. We will detail our approach highlighting the link between the definition and usefulness of contextual information. We will introduce our context definition and we will show how contextual elements are constructed based on the proposed definition. At the end of section III, we will propose some scenarios to demonstrate the consistency of our definition. Section IV concludes the paper with a discussion, our contributions and presents our future work.
.

## II. RELATED WORK

So far, the context has been defined as the set of circumstances or facts that surround a particular event or situation, which is the essence of the definitions proposed in several researches [8], [2], [12], and [28].

The next subsection enumerates the most significant trends we noticed in context definitions.

### A. First Trend: Physical Aspect

Most the first researches on the context, focused on a physical enumeration of contextual factors (element/ data).

In their work Schilit et al [5], have considered that the context has three important aspects related to the following questions: Where are you? With whom? What resources do you have nearby? They defined context as changes in the physical environment using: location, description of people and objects in the immediate environment and changes in these objects. Brown restricted the context to elements of the user environment, "the elements of the environment that the user's computer knows about" [24]. Later, Brown, Bovey and Chen, introduced the season, the temperature, the identity and location of the user to the context definition [25]. They presented a set of extensible components to characterize the context in which the basic elements are: the location, all items, the user needs, time, and spatial orientation.

While we agree with most of these definitions (in particular the physical aspect of context elements), we find they did not take into account logical and abstract data (e.g. time), resulting in changes in the system state and services.

### B. Second Trend: Towards Time Aspect

Rayan defined the context element: "the elements of context are: user location, environment, identity and time." [20]. Ward saw context as statements of probable application environments [4], "states the possible application environments". Pascoce described the context as a subset of physical and conceptual states of interest to a particular entity [11]. Schmidt defined the context as "knowledge about the user and the statements of equipment, environment and location." [3]. While Brézillon "perceived it as everything not explicitly involved in solving a problem, but constrains it" [22].

In these definitions, time is implicitly or vaguely defined. However, time is a crucial element of context. It is also an element that should be defined according to the purpose of the application. Based in purpose, time information may change. As we will show in the sub-section III-a. and III-c.

### C. Third Trend: Adaptation-Driven Context Definition

In their study of context awareness in mobile, Chen and Kotz showed that general context definitions remain vague and inadequate in a computing environment [8]. They defined the context as the set of environmental states and parameters that determine the behavior of an application or in which an event of the application runs and is of interest to the user. Thus research on context definition was oriented towards adaptation.

Dey emphasized the relevance of information by providing a definition where he attempts to clarify the nature of the entities of the context:" Context is any information that can be used to characterise the situation of an entity. An entity is a person, place, or object that is considered relevant to the interaction between a user and an application, including the user and applications themselves." [1].

This definition encapsulates all the above definitions since it is very generic. Dey reported that the context parameters can be implicit or explicit. Indeed, the context parameters may be provided by the user or by sensors located in the environment of the user and the application. They can also be reached by a more or less complex interpretation of these parameters. Dey gave a definition covering any implicit or explicit data that can be critical to the application. In 2001, Winograd approved the definition given by Dey and argued it covers all existing work on context. However, he considers that the expressions used by Dey such as "any information" and "characterize an entity" very general and not marking any limit to the concept of context (anything could be a context). In order to bring more precision to the definition of Dey, Winograd presented context as a set of structured and shared information. He explained this definition by stating "the consideration of information as context is due to how it is used and its inherent properties." He supported his idea by the following example "The voltage on the power lines is context if there is some action by the user and/or computer whose interpretation is dependent on it, but otherwise is just part of the environment." [31]. In 2002, Henricksen defined context as "the circumstance or situation in which a computing task takes place." [12].

An analysis by Brezillon [23] conducted on the definitions of context led to the conclusion that most definitions are answers to the following questions: Who? To identify the current user and others in the environment; what? Where? When? Why? How? To describe how the activity takes place. Answering these questions can generate a large set of information, most of it is unnecessary in addition to management problems of this information and storage in the context of mobile systems. In 2005, Truong gave a new definition of context based on Dey's, but put more emphasis on the role of context in interaction. He considered the context as a set of directions and information that define the interaction between the user and the application [5].

In a more abstract version without detailing the components of the context, Miraoui described the context as "Any information that trigger a service or change the quality of a service if its value changes"[18]. He explained, "This definition is a compromise between the abstraction of the term context and the limitation of the set of contextual information related to services provided (goal of pervasive computing). It does not enumerate examples of context which makes it more generic and independent of the application. It does not take into account information that might characterize context but do not play a relevant part in service adaptation [18].

According to Gensel the context is the set of characteristics of the physical or virtual environment that affects the behavior of an application and whose representation and acquisition are essential for the adaptation of information and services [10]. For Najar, the context acts as an external element to the computer system that affects its internal variability, and as a parameter guiding the variant and the process of adaptation [28].

Some of those definitions remain vague and general such as Dey's definition, or neglected the time information, which is important in the prediction, or provided information that is not necessary for the application.

*D. Towards Adaptation and Prediction Trends*

Up to now, most of the work on context-aware adaption services focused on the current context. Recently, a new trend supported the idea of taking into account the temporal evolution of context to predict a future context. A noteworthy example is the work proposed by [26], [27], [17].

While these researchers have demonstrated the importance of the use of prediction and future context, they did not elaborate the context definition or the purpose of its use.

Although, they all had to revisit the old definitions, trying to identify the contextual information they would use in their modeling during the service adaptation or prediction according to user behavior and as part of a specific application.

As a result we did not find a specific definition adapted to prediction, especially in a general and reusable form.

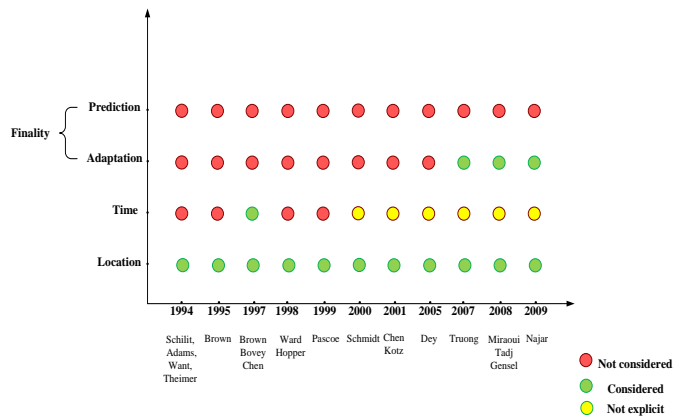

Figure 1: Comparative graphical of context definitions

*E. Synthesis*

Definitions of context remain general, vague and do not reflect the new trend in service adaptation based on prediction or context discovery.

All the recent work showed the importance of being aware of the future context, which allows a pervasive system to choose the most effective strategies to achieve its objectives and therefore makes it possible to provide an active and fast adaptation to future situations.

We conclude with a comparative diagram shown in Figure.1, summarizing the overall trends of the context definitions while putting forward the presence of the criteria that we consider crucial in defining a context: location, time and purpose. We are therefore interested in modeling the context in such a way that it takes into account the prediction or/and adaptation.

### III. PROPOSED DEFINITION

In the following, we identify three essential lines that fall within the definition of context. These lines stem from our own approach to define the context.

*A. Link Context and Time*

The context must be defined in accordance with the space-time domain. This observation is consistent with Ryan's point of view with respect to the importance of time in the characterization of the context [20]. In fact, any situation, event or interaction should be performed within a mandatory interval of time. The granularity of this interval is crucial since it presents the temporal extension of the information describing the context. This fact has also an impact on the quantity of information: one can pass from a small number of information entities to a database, depending on whether one focuses on an event of short duration or a sequence of events (activity).

We also strengthen our approach with the affirmation of Jensen [7]. He distinguishes two aspects of time that are very specific in the description of temporal data:

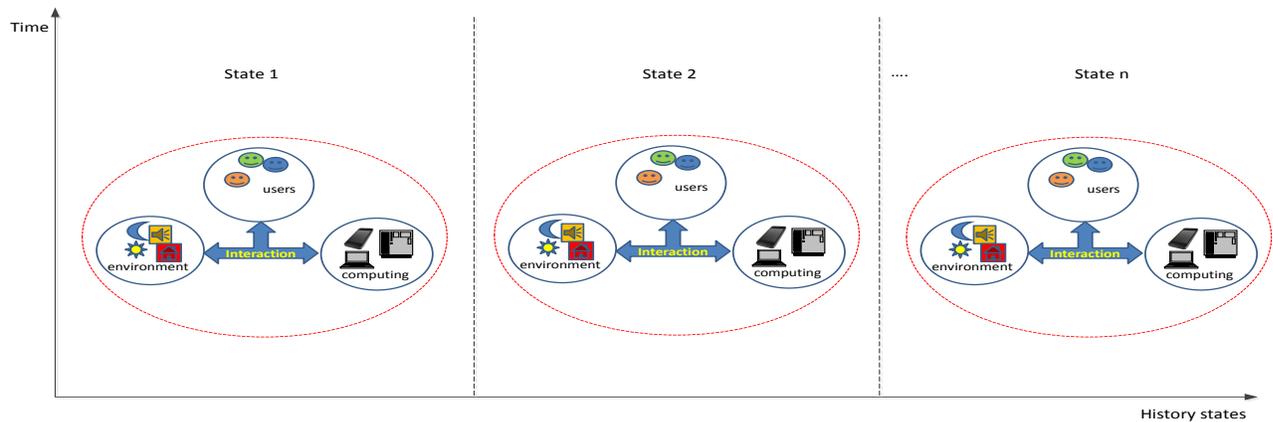

Figure 2. Conceptual diagram in the context of space-time

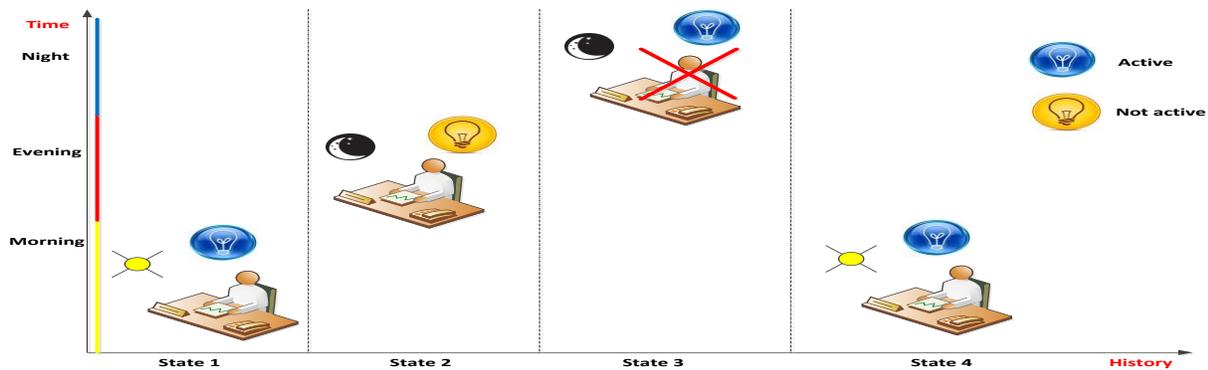

Figure 3: Example smart Luminosity control in an office according to user presence and time.

(i) the valid time (the time, the phenomenon modeled in reality was produced) and

(ii) The transaction time (the time, data on this phenomenon was introduced in the database).

We conclude then that if the time isn't defined during the production of the event or introduced as an explicit element of the context, we would have estimation failures in time during prediction, which affect the service quality.

Therefore, any entity (user, physical or logical environment component, computer component) used in the interaction around an event in a pervasive system must be considered in relation to a time axis describing its dynamics. This helps us recognize a prior state (past), a current state (present) or a future state, to be estimated following a temporal logic, as shown in Figure.2 and in Figure.3.

### B. Link Context and Space

In the literature, the spatial context is reduced, in the best case, to a variable location (user location). As stated by M. Georgia et al "We can identify the places located in the relative user environment." [15].

We can rather find the concept of "context usage" that defines and includes any information characterizing the user, the environment and the platform [19].

Li [14], presented a definition of spatial context (or geographical context) in three dimensions: (1) the static spatial context, (2) the dynamic spatial context and (3) the internal spatial context.

He defined them as follows:
• The static spatial context: is geo-spatial information that may have an impact on the environment of the user. For example information about the road network, buildings, shops, interior structures (rooms, offices, flats ...).
• Dynamic spatial context: gathers geo-referenced information collected by different types of sensors. For example, the speed data collected on the roads, the data on the occupancy rate of parking lots, etc.

• The internal spatial context: gathers information available from local devices (e.g. GPS). For example, a motorist may have information about his position, speed, direction, destination with his GPS. He may also obtain information from an integrated sensor such as a vehicle's fuel gauge.

### C. Relevance of Context According to the Purpose of Use

O.Bucur [21] deepened the characterization and formulation of the relevance of context information by integrating the concept of finality context (define contextual elements according to the purpose of their use). The finality refers to the purpose for which the context is used at a given time. This purpose can be short or long term.

If the purpose is set to long term, the context is archived over time while maintaining relevance. For example, if a system is designed to adapt the user interface according to usage patterns, then the data of past uses continue to be relevant because they can determine the next adaptations and predict the future actions of the user. If purpose is short-term (instantaneous) the current context will be used immediately in the process of adaptation. The example is the transition to the silent mode of the user's mobile phone upon entering the library.

We conclude, these questions regarding: time, spatial context, and purpose, still present a major challenge in the design of systems, applications and context-aware services. We discuss this in the following section with more details.

*D. Proposed Context Definition*

Our definition promotes following axes: (1) presenting the time as a most important feature of context, (2) redefining the spatial dimension in the environmental integrity of the system and services in a usage context, (3) and incorporating the finality of the context in the definition of contextual information. This means that, the elements of context should be considered according to the objective for which the context is used. We notice then, that the prediction oriented definition is mandatory. Knowing the adaptation scenario based on the current context does not require the identification of the spatiotemporal or temporal information, neither the finality of the use of context. However, the attributes are crucial for prediction.

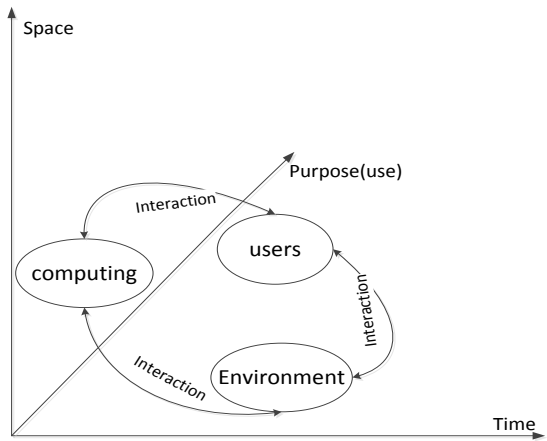

Figure 4. Position of a context according to the 3 axes (time-space-purpose)

From our point of view, the context in a pervasive system is located around a spatiotemporal variation and is defined by the three axes mentioned above in Figure 4: (1) Time (2) Space, and (3) Purpose (finality) of its use. Therefore, we summarize our context as: "**Any entity undergoing a spatiotemporal variation and that may lead to a change in the service or the quality of service in the short or long term**".

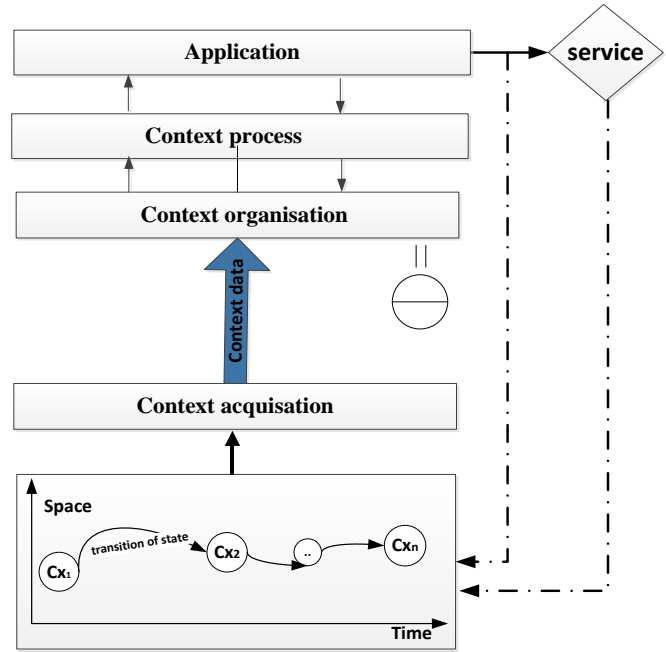

Figure 5. Context awareness operations

This definition provides a description of a context able to answer: Where? When? What for? And any other information for which the change of the value triggers a service or change the quality of a current or future service. Thereby, it can be useful in an habitual or exceptional system behavior (e.g. failure sensor, user habit change).

Where? As well as the geographical location, this also describes the spatial location of the use in a dynamic, static and internal dimension,

When? The most real time temporal data a real-temporal data with consideration of possible granularity of time measurement in a dynamic and proactive system such as in pervasive systems

What for? To understand the finality behind the information as defined as an element of context and determine the purpose of the use of the contextual information for the short and long term.

As shown previously, the context is following a spatiotemporal dynamic, that needs to be taken into account, when acquiring contextual data (as mentioned in our definition). Once the data acquired, we organize it, according to its projected use to make it easily available to the relevant processes.

The context generated at the "process" step, which might be either the current context or the predicted context, will influence the application's behavior. Conversely, once the application updates itself after new adaptations triggered by underlying services, the context will in turn be updated.

We believe that this definition identifies the most relevant contextual information and is directly involved in the design of a context-aware system. It could be oriented adaptation and prediction (Figure 5).

*E. Demonstration Scenarios*

The following scenarios are presented to show the consistency of the proposed definition.

*e.i. Mobile Scenarios: Screen Luminosity Control System*

A screen of a cell phone increases its luminosity when it is in a dark place and reduces when it is in lighted area, as shown in Figure 6.

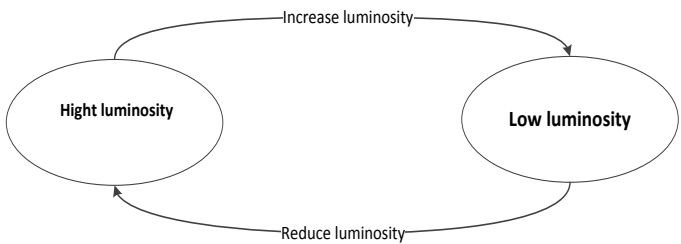

Figure 6. State diagram

In traditional scenario, the service will change depending on the information defined by a brightness sensor.

Applying previous definition oriented adaptation that gives the following description of service, service form, and context information.

- **Service:** luminosity adjustment of the screen.
- **Service form:** increased or reduce luminosity
- **Context information:** - Lighting

Table 1. Contextual information scenario (1)

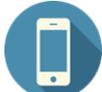

| Equipment | service | Form | Contextual information | Contextual information for prediction | Contextual information for adaptation |
|---|---|---|---|---|---|
| Smart-phone | Luminosity Adjustment | Increase luminosity | Lighting | Ddiscover context (prediction) | Selection context (adaptation) |
| | | | Location | Location | Lighting |
| | | Reduce luminosity | Time | Time | |
| | | | Previous state | Previous state | |

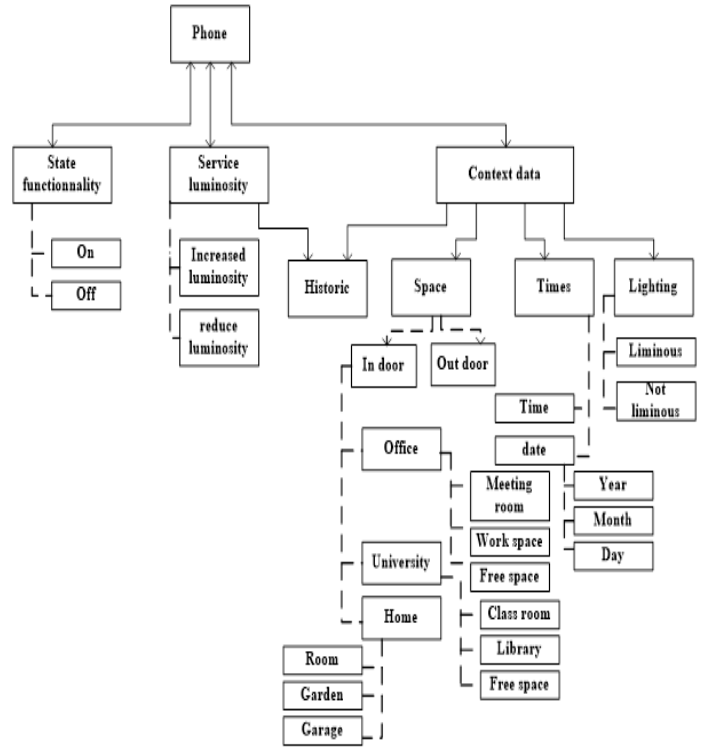

Figure 7. Model of luminosity adjustment service

If brightness sensor fails, the context prediction can provide a solution to ensure the continuity of service.

Based on previous definitions we cannot have solution to make decision about service, and we cannot specify the information, according to the adaptation module or prediction.

Based on our definition, detecting contextual information will be easier, more accurate, and specific for each module (prediction, adaptation) as shown in Table.1 and following description (Figure 7).

- **Service**: luminosity adjustment of the screen.
- **Service form**: increased or reduce luminosity
- **Context information**: - Lighting - Location - Time Previous state.

*e.ii. Home Automation Scenarios: Home Heating Control System*

For energy-saving heating should be turned off when the user leaves home in the morning for work. The user is primed to leave work and go home, and he wants to have his home in an ambient temperature when he returns there.

[A] heater                    [1] home heating control system environment
[B] thermostat                [2] global user's environment
[C] motion detector
[D] smart home control system

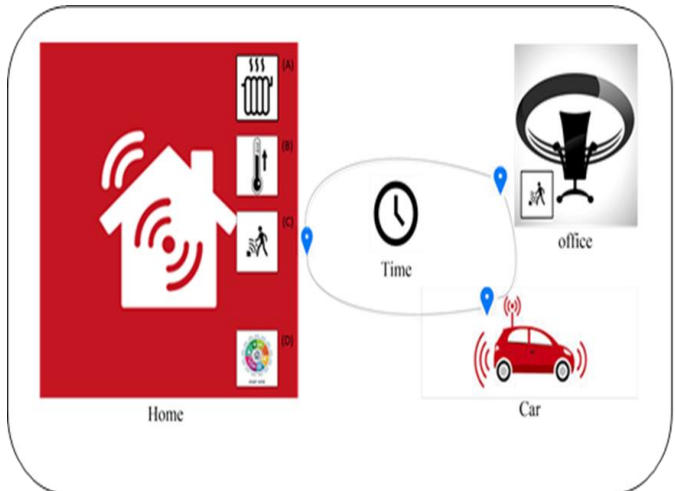

Figure 8. User's overall environment

In the traditional adaptation process, the heating system is activated when it detects the user presence in the house. Achieving the desired temperature takes a little time, (Figure 8: (1) home heating control system environment).

Whereas in an adaptation process based on prediction, the system starts preparing the home when the user is still on the way, to get the desired temperature by the time the user will just reach home. By estimating the time according to itinerary indicated by the user's car GPS, the service system can trigger heating, sometime before the arrival of the user at home (Figure 8: (2) home heating control system environment).

Applying previous definition oriented adaptation that gives the following description of service, service form, and context information.

- **Service**: heating temperature adjustment in house.
- **Service form**: on or off heating
- **Context information**: - User intern location
                          - Actual temperature

Providing the service before physical presence of user in home, autonomously, proactively and invisibly remains impossible. Service will depend on a manual user command.

Based on our definition, detecting contextual information will be easier, and specific for each module (prediction, adaptation) as shown in table.2 and following description (Figure.9). It makes possible the autonomous heating service without manual user command or physical user presence.

- **Service:** heating temperature adjustment in house.
- **Service form:** on or off heating
- **Context information:** -User intern location
                          -Actual temperature
                          -User external location
                          -Time

The following table shows all the contextual information necessary in scenario (2) according to the purpose.

Table 2. Contextual information scenario (2)

| service | Contextual information | Contextual information for prediction | Contextual information for adaptation |
| --- | --- | --- | --- |
| Heating temperature adjustment | User intern location | Actual temperature | User intern location |
|  | Actual temperature | User extern location | Actual temperature |
|  | User extern location | Time |  |
|  | Time |  |  |

The overall conclusion, our definition offers an efficient approach to define contextual information in a pervasive system. It establishes clear parameters that define efficiently the context. It takes into account the dynamism of a pervasive system. It encompasses adaptation and prediction takes into account the purpose giving a more efficient design. And allows exceeding the previous definitions limits, to define useful contextual information for prediction and proactive adaptation.

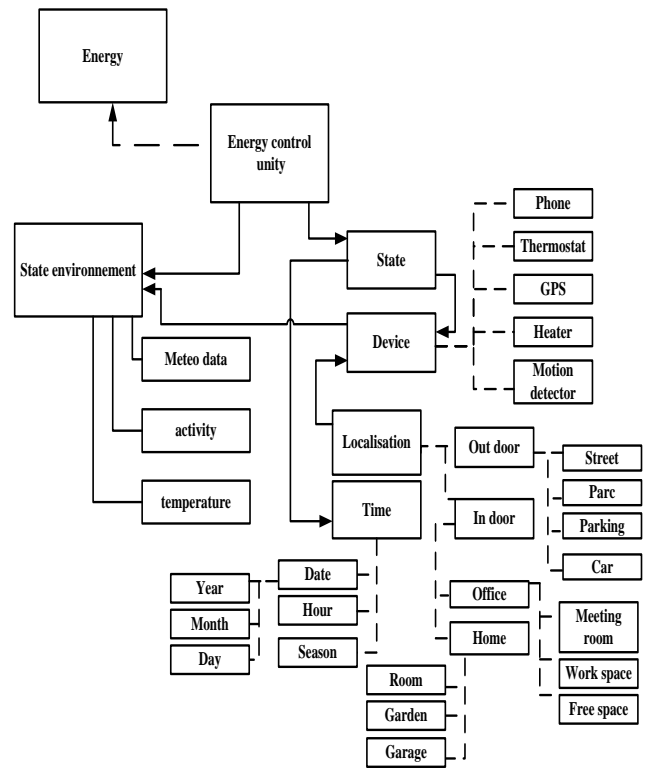

Figure 9. Model of heating adjustment service

IV. CONCLUSION AND FUTURE WORK

In this paper, we presented a new definition of the context, taking into account 1) the spatiotemporal information, 2) the purpose of use, and 3) any other information that may be useful and leads to a current or future change in the adaptation of service. We presented an overview of works that have discussed the definition of context and its adaptation to the service, and we proposed a new vision of the definition of context. In the scenarios proposed in the section III, we showed the difference in the contextual information oriented adaption or oriented prediction, which proves our vision.

Future work will concentrate mainly, on context modeling and prediction method. Taking into account the complexity of identifying the elements of the context in a pervasive system, in our future work, we will propose a method to model contextual information. We are going to develop architectural model of context-aware systems incorporating the prediction approach. Currently we focus to create our prediction approach based on our spatiotemporal context vision.